\documentclass[aps,pre,superscriptaddress,twocolumn,balancelastpage]{revtex4-1}

\usepackage[colorlinks,bookmarks=false,citecolor=blue,linkcolor=blue,urlcolor=blue]{hyperref}
\usepackage[all]{hypcap}   % let hyperlinks correctly point to figures rather than their captions;

\usepackage{amsmath,amssymb}
\usepackage{graphicx}
\graphicspath{{figures/}{/}}
\usepackage{verbatim}
\usepackage{color}

\usepackage{placeins}  % for FloatBarrier
\usepackage{flafter}     % Bilder immer nach \figure Befehl

\usepackage{physics}

\usepackage{color}

\newcommand{\ue}{\text{e}}
\newcommand{\ui}{\text{i}}
\newcommand{\ud}{\text{d}}
\newcommand{\suba}{{\text{1}}}
\newcommand{\subb}{{\text{2}}}
\newcommand{\Jxa}{J_x^{(\suba)}}

\newcommand{\Jza}{J_z^{(\suba)}}
\newcommand{\Jxb}{J_x^{(\subb)}}

\newcommand{\Jzb}{J_z^{(\subb)}}

\newcommand{\xa}{x_{\suba}}
\newcommand{\ya}{y_{\suba}}
\newcommand{\za}{z_{\suba}}
\newcommand{\xb}{x_{\subb}}
\newcommand{\yb}{y_{\subb}}
\newcommand{\zb}{z_{\subb}}

\newcommand{\coupling}{\epsilon}
\newcommand{\cherrycoupling}{\mu}
\newcommand{\crit}{\text{crit}}

\newcommand{\mpstate}{\ket{\leftrightarrows}}
\newcommand{\pmstate}{\ket{\rightleftarrows}}

\usepackage[normalem]{ulem}

\newcommand{\HIDDEN}[1]{}

\begin{document}

\title{Fast Bit-Flipping based on a Stability Transition of Coupled Spins}

\author{Maximilian F.~I. Kieler}
\affiliation{Technische Universit\"at Dresden,
 Institut f\"ur Theoretische Physik and Center for Dynamics,
 01062 Dresden, Germany}

\author{Arnd B\"acker}
\affiliation{Technische Universit\"at Dresden,
 Institut f\"ur Theoretische Physik and Center for Dynamics,
 01062 Dresden, Germany}

\date{\today}

\begin{abstract}
A bipartite spin system is proposed for which a fast transfer from one defined
state into another exists.
For sufficient coupling between the spins, this implements a
bit-flipping mechanism which is much faster than that induced by tunneling.
The states correspond in the semiclassical limit to equilibrium points
with a stability transition from elliptic-elliptic
stability to complex instability for increased coupling.
The fast transfer is due to the spiraling characteristics of the
complex unstable dynamics. Based on the classical
system we find an approximate scaling relation for the
transfer time, which even applies in the deep quantum regime.
By investigating a simple model system, we show
that the classical stability transition is reflected in a
fundamental change of the structure of the eigenfunctions.
\end{abstract}

\maketitle

The concept of quantum computing \cite{NieChu2010} enables new
possibilities of future computational devices. Despite a lot of recent progress,
quantum computing will not supersede classical computing, instead they are
expected to complement each other
\cite{PerMcCShaYunZhoLovAspBri2014,McCRomBabAsp2016}.
Thus it is important to investigate current quantum computation
realizations for the ability of implementing classical operations.
A fundamental requirement of information processing is a fast and energy
efficient switching between two (quantum) states $\ket{0}$ and $\ket{1}$.
From a many-body perspective, this could be realized for example by using
spintronics \cite{BhaSbiHirOhnFukPir2017,BalManTsoMorOnoTse2018,ZutFabDas2004}.
Another approach is to employ few-body systems, which should be realizable
with current experimental devices, e.g.\ using ultra cold atoms
\cite{BloDalNas2012, AlbGatFoeHunCriObe2005,TomMueStrLoeSchKetObe2017}.
Such many body systems differ from spin $\frac{1}{2}$ systems by having a
semiclassical limit and thus can be investigated by more intuitive
classical methods.
For example, a transfer between two specific states can be realized using
dynamical tunneling \cite{DavHel1981,KesSch2011,SteOskRai2001, Hel2018} between
symmetry related regions. However, the tunneling time for such a quantum
process approximately depends inversely on the energy difference of the two
states, hence demanding for a compromise between energy efficiency and switching
speed.
Thus it is an interesting open question to devise systems for which both goals
can be reached simultaneously.

\begin{figure}[b]
    \begin{center}
        \includegraphics[width=8.6cm]{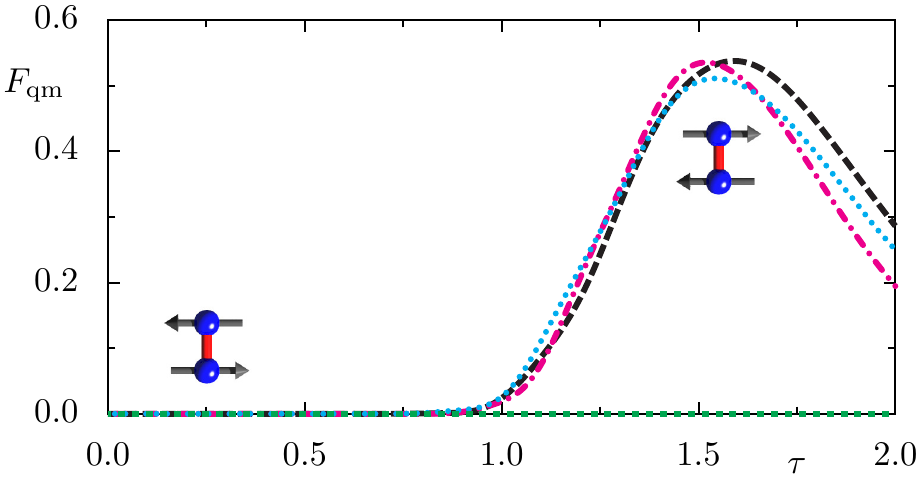}
    \end{center}
    \caption{Quantum transfer
            from $\ket{0} \equiv \pmstate$ to $\ket{1}\equiv\pmstate$
            as  measured by the fidelity \eqref{eq:quant_transfer}
            for different parameters $(j, k, \epsilon)$ as $(25, 1.0, 1.2)$ for
            black dashed, $(40, 1.0, 1.2)$ for magenta dash-dotted and
            $(25, 1.2, 1.4)$ for cyan dotted line. The
            transfer for different parameters follows the same curve by
            using the re-scaled time $\tau$, see Eq.~\eqref{eq:rescaled_time},
            which is the actual time divided by the time of the analytical
            approximation, explained in more detail in the
            main text below.
            The green dashed curve with $F_{\text{qm}} \approx 0$
            is for $(j, k, \epsilon) = (15, 1.0, 0.8)$ where only dynamical
            tunneling occurs. The fast bit-flipping for the first sets of
            parameters semiclassically corresponds to complex unstable
            dynamics (defined below) while no transfer occurs for
            elliptic-elliptic stability.}
        \label{fig:fig_cu_transfer}
\end{figure}

In this paper we present a mechanism for bit-flipping
in a system of two coupled spins, which is both fast and energy efficient.
We implement the states representing the bits by anti-parallel aligned
coherent states \cite{Per1986, ZhaFenGil1990} of the spin system, i.e.~$\ket{0}
\equiv \pmstate, \ket{1} \equiv \mpstate$. In the classical limit of large spins, the
quantum system becomes a dynamical system and the states can be classically
described by initial conditions which are localized around the equilibrium
points of the system.
Depending on the parameters the stability of these equilibrium points
changes from elliptic-elliptic (EE) stability to complex instability (CU)
\cite{HowMac1987a,Sko2001b}.
The characteristic feature of a complex unstable equilibrium
is a spiraling dynamics in the neighborhood leading to a fast repulsion away
from the equilibrium point \cite{Heg1985, ConFarPapPol1994, StoBae2021,
PatManChaSkoPue2022}.
As the dynamics is confined to the surface of Bloch spheres for each
subsystem the spiraling motion transfers the state from one side to the
anti-podal side. This classical transfer is also reflected
in the quantum time evolution of the state $ \pmstate$
which transforms into the bit-flipped state $ \mpstate$.
The transfer to $ \mpstate$ can be quantified
using the fidelity \cite{Joz1994}, see  Fig.~\ref{fig:fig_cu_transfer}.
We establish based on classical arguments a universal behavior
in system size using a re-scaled time.
Using a simplified system, which embodies the essential
aspects of the  stability transition, we demonstrate
that this is connected with a fundamental change
in the structure of the eigenfunctions from localized to non-local.

%%%%%%%%%%%%%%%%%%%%%%%%%%%%%%%%%%%%%%%%%%%%%%%%%%%%%%%%%%%%%%%%%%%%%%%%%%%%%
\emph{Quantum system}\label{seq:the_quantum_system} ---
We consider a bipartite system of two coupled spins
\begin{equation}
 \begin{aligned} \label{eq:coupled_top_hamiltonian}
  H &=    \Jxa + \frac{k_\suba}{2 j_\suba} \left( \Jza \right)^2
        + \Jxb + \frac{k_\subb}{2 j_\subb} \left( \Jzb \right)^2 \\
        &\qquad + \frac{\coupling}{\sqrt{j_\suba j_\subb}} \Jza \Jzb,
 \end{aligned}
\end{equation}
where $J^{(i)} = (J_x^{(i)}, J_y^{(i)}, J_z^{(i)})$ is the spin operator of the
subsystems $i=1, 2$. For simplicity we restrict ourselves to spins of
equal size, $j_\suba = j_\subb = j$, giving rise to
subsystem Hilbert spaces of dimension $N = 2 j + 1$. For
numerical illustrations $k_\suba = k_\subb = k = 1.0$ is used
(unless explicitly mentioned).
The parameter $\coupling$ imposes a coupling of the two subsystems, each being
a variant of the Lipkin-Meshkov-Glick model \cite{LipMesGli1965}.
Note that the system \eqref{eq:coupled_top_hamiltonian} is an autonomous
version of the coupled kicked tops \cite{MilSar1999a,BanLak2004}. It can also
be written in terms of Bose-Hubbard operators
\cite{StrGraKor2008, MicJakCirZol2003} and experimentally be realized in its
time-periodically driven version \cite{TomMueStrLoeSchKetObe2017}.

The bit-states $\ket{0}$ and $\ket{1}$ are realized by two spin-coherent
product states \cite{Per1986, ZhaFenGil1990, StrGraKor2008}. The first state,
$\ket{0} \equiv \pmstate$, aligns both angular momenta anti-parallel in
$x$-direction, i.e.,~the first spin in positive and the second in negative
$x$-direction, see Fig.~\ref{fig:classical_motion}. The second
state $\ket{1} \equiv \mpstate$ is the spin-flipped counterpart.
The time evolution is given by the unitary time evolution operator
$U(t) = \ue^{-\ui H t}$ and we quantify the quantum transfer from $\ket{0}$ to
$\ket{1}$ by the fidelity between the time evolved state
$\ket{\leftrightarrows(t)} = U(t) \ket{\leftrightarrows}$ and the
fixed bit-flipped state $\mpstate$,
\begin{equation} \label{eq:quant_transfer}
 F_\text{qm}(t) = |\braket{\rightleftarrows}{\leftrightarrows(t)}|^2.
\end{equation}
Therefore $F_\text{qm}(t=0) = 0$ and a full transfer to the bit-flipped
state would correspond to $F_\text{qm}(t_\text{trans})=1$.
It turns out that there is a parameter range of the coupling $\coupling$
for which there is essentially no transfer.
However, increasing the coupling beyond some critical parameter
$\coupling_{\crit}$, see Eq.~\eqref{eq:transition_boundary} below, a fast and
significant transfer takes place, illustrated in
Fig.~\ref{fig:fig_cu_transfer}. We point out that this is not a full transfer
of the state, but it is sufficient to be reliably detectable.
In the following, we explain the mechanism underlying this transfer.

\begin{figure}
    \begin{center}
        \includegraphics[width=8cm]{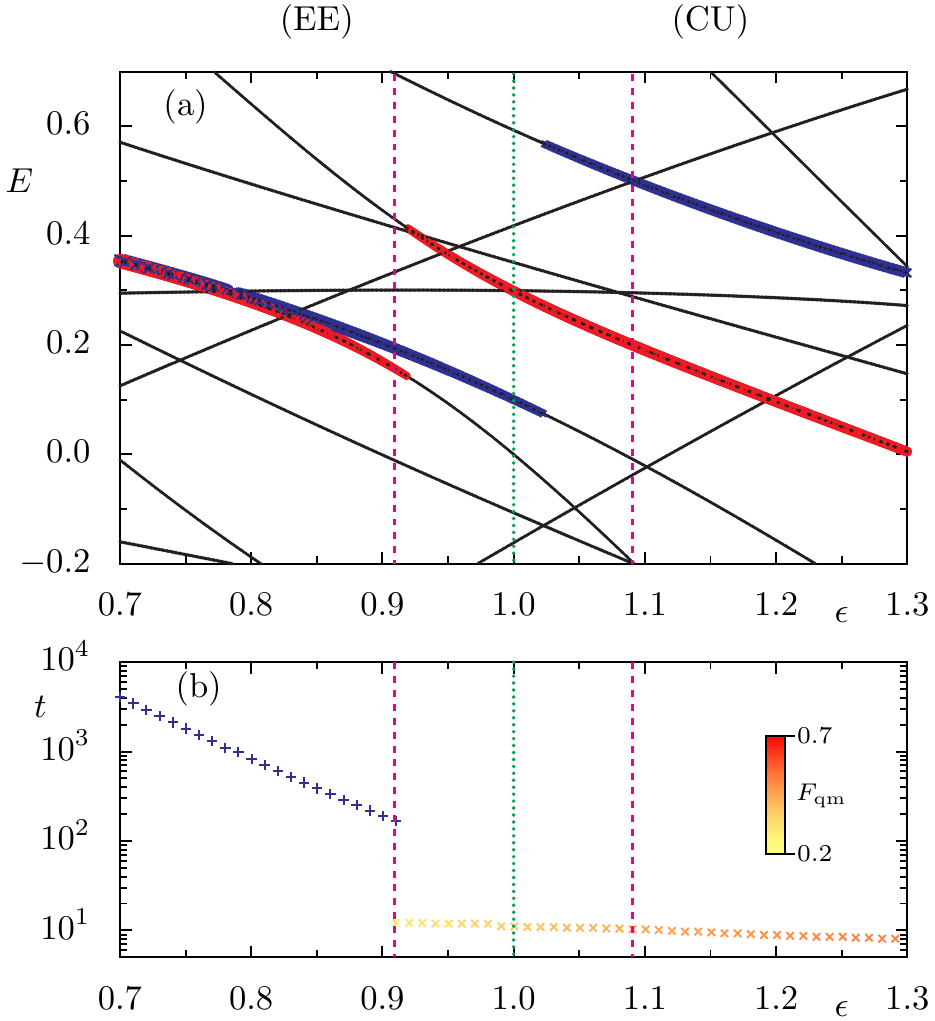}\\
        \caption{(a) Relevant energy levels as function of the coupling
                 $\coupling$: Red lines highlight those
                 levels whose eigenfunction have
                 maximal overlap with $\pmstate$ and
                 blue lines those with $\mpstate$.
(b) Dynamical tunneling time $t_\text{dyn}$ (blue $+$) for $\coupling <
\coupling_\crit=1$
and the transfer time $t_\text{trans}$ (yellow to orange $\cross$) for
$\coupling > \coupling_\crit$, see
Eq.~\eqref{eq:transition_boundary}, in semi-logarithmic scale.
The fidelity at time $t_\text{trans}$ is encoded by color.
The green vertical dotted line indicates the critical coupling
strength
$\coupling_\crit = 1.0$ and the red vertical dashed lines the parameter regime
$\coupling_\crit \mp 1/N$ for $N=11$ ($j=5$).
}
        \label{fig:energy_level_plot}
    \end{center}
\end{figure}

The quantum dynamics of the bipartite quantum system
\eqref{eq:coupled_top_hamiltonian}
is determined by the eigenvalue equation $H \ket{\Psi_n} = E_n \ket{\Psi_n}$.
Therefore, a representation of the energy levels $E_n$
as function of the coupling $\coupling$, for a small system, $j=5$,
see Fig.~\ref{fig:energy_level_plot}, provides an intuitive understanding
of the mechanism:
For the time-evolution, $U(t) = \sum_n \ue^{-\ui E_n t} \dyad{\Psi_n}$,
only those eigenstates are relevant which are related to the states $\pmstate$
and $\mpstate$.
Due to the anti-parallel structure of the spins in $\pmstate$ and $\mpstate$,
the sum of the energies of the identical subsystems is approximately zero
so that we can concentrate on the middle of the spectrum only.
The eigenstates, which are most relevant for the transfer from $\pmstate$ to
$\mpstate$ can be identified by computing the overlap of $\pmstate$ and
$\mpstate$ with all eigenstates, respectively,
e.g.~$|\braket{\leftrightarrows}{\Psi_n} |^2$. In
Fig.~\ref{fig:energy_level_plot}(a), we present the energy levels whose
eigenfunctions have the maximal overlap with $\pmstate$ in blue and with
$\mpstate$ in red.
Firstly, we focus on the parameter regime $\coupling < k$ for which the
overlap of $\mpstate$ and $\pmstate$ occurs mainly with two eigenstates only.
In this parameter regime the two states are in a semiclassical description (see
below) located within symmetry-related regular regions in phase space, so that
dynamical tunneling \cite{DavHel1981,KesSch2011,SteOskRai2001, Hel2018} between
them is possible. This  situation can be described by an
effective two level system with eigenfunctions of the form $\ket{\Psi_{\pm}} :=
\frac{1}{\sqrt{2}} \left(\pmstate \pm \mpstate \right)$. The tunneling time is
determined by $t_\text{dyn} = 2 \pi / |E_{+} - E_{-}|$. However, as the
considered pair of energy levels is nearly degenerate, the tunneling time
becomes very large, see Fig.~\ref{fig:energy_level_plot} (b),
hence a transfer from $\pmstate$ to $\mpstate$ by dynamical tunneling would
require extremely long times.
This is seen in Fig.~\ref{fig:fig_cu_transfer}, where the fidelity
remains essentially zero in the considered time interval.

By increasing the coupling to $\coupling \sim k$, the system starts to behave
differently. The state distributes over multiple eigenfunctions
with larger energy gaps and the approximation by a two-level system becomes
invalid. Hence the dynamical tunneling time does not provide a
suitable estimate in the transition regime. For the parameter regime $\epsilon
\gtrsim k$ the transfer time $t_\text{trans} =
\max(F_\text{qm})$, i.e.\~the time to reach the first maximum, is
numerically computed using the fidelity, Eq.~\eqref{eq:quant_transfer}.
This transition occurs much faster than for dynamical tunneling, see
Fig.~\ref{fig:energy_level_plot}(b). Note that the dynamical
tunneling time is the time to reach the first global maximum and is therefore
different from the transfer time, which is the time to reach the first
maximum. In Fig.~\ref{fig:energy_level_plot} the fidelity achieved at the
transfer time is indicated by color. This shows, that for having both
short transfer time and large fidelity one needs  $\epsilon > 1$.
Moreover, as the initial and final state still only have a small energy
difference, the transfer can be done with less energy than for example in a
single spin system, where the energy difference between opposite spin
configurations is much larger. While the dynamical tunneling for $\coupling <
k$ is a purely quantum effect, the transfer for $\coupling > k$ is of classical
origin and can be approached by semiclassical methods.

%%%%%%%%%%%%%%%%%%%%%%%%%%%%%%%%%%%%%%%%%%%%%%%%%%%%%%%%%%%%%%%%%%%%%%%%%%%%%
\emph{Semiclassical description} --- By considering the semiclassical limit of
the quantum spin to a classical angular momentum leads to
a dynamical system. Using the mean-field approach,
see e.g.\ Ref.~\cite{StrGraKor2008},
of replacing operators by c-numbers and taking the limit $j \to \infty$ we
obtain from Eq.~\eqref{eq:coupled_top_hamiltonian}
the system of differential equations
\begin{equation}
\begin{aligned} \label{eq:cl_coupled_top_ode}
 \dot \xa &= -k_\suba \ya \za - \coupling \ya \zb  \\
 \dot \ya &= -\za + k_\suba \xa \za + \coupling \xa \zb  \\
 \dot \za &= \ya \\
 \dot \xb &= -k_\subb \yb \zb - \coupling \yb \za \\
 \dot \yb &= -\zb + k_\subb \xb \zb + \coupling \xb \za \\
 \dot \zb &= \yb.
\end{aligned}
\end{equation}
The pairs of coordinates $(x_i, y_i, z_i)$, for $i = 1,2$, each lie on the
surface of a unit Bloch sphere. Thus the system \eqref{eq:cl_coupled_top_ode}
can be mapped by a canonical transformation $(\phi_i, z_i) =
(\sqrt{1 - z_i^2}\arctan(y_i/x_i), z_i)$ into a system which is effectively
four dimensional with coordinates $(\phi_1, z_1, \phi_2, z_2)$.
In this classical description, the coherent states $\pmstate$ and
$\mpstate$ turn into equilibrium points of \eqref{eq:cl_coupled_top_ode},
i.e.,~$\pmstate$ corresponds to
$(x_1, y_1, z_1; x_2, y_2, z_2) = (1, 0, 0; -1, 0, 0)$
and $\mpstate$ to $(-1, 0, 0; 1, 0, 0)$, respectively.
Thus the quantum transfer from $\pmstate$ to  $\mpstate$
can be investigated in terms of the classical dynamics
of orbits in the neighborhood of these equilibrium points.
The stability of such orbits is determined by the linearized dynamics which is
characterized by the four stability eigenvalues $\lambda_\ell$ of the stability
matrix \cite{HowMac1987a}.
Depending on the system parameters a transition from elliptic--elliptic
(EE) stability to complex unstable (CU) dynamics occurs for the critical
parameter
\begin{equation} \label{eq:transition_boundary}
    \coupling_\crit = \frac{k_\suba + k_\subb}{2}.
\end{equation}
The elliptic stability for $\coupling < \coupling_\crit$ is characterized by
purely imaginary eigenvalues. As the local dynamics effectively corresponds to
$\ue^{\lambda_\ell t}$, the dynamics is rotational and stays in a bounded
neighborhood of the equilibrium, hence classically no spin flip is possible. In
the CU parameter regime, $\coupling >
\coupling_\crit$, the four eigenvalues $\lambda_\ell$
form a so-called Krein-Quartet
$(\pm \lambda, \pm \lambda^*)$ with some complex $\lambda =  c_1
 + \ui c_2$. This leads to a logarithmic spiraling motion~\cite{Heg1985,
ConFarPapPol1994, StoBae2021, PatManChaSkoPue2022}, see
Fig.~\ref{fig:classical_motion}, where the expanding motion is determined by
the real part $c_1$ and the rotation by $c_2$.
\begin{figure}
    \begin{center}
        \includegraphics[width=8.6cm]{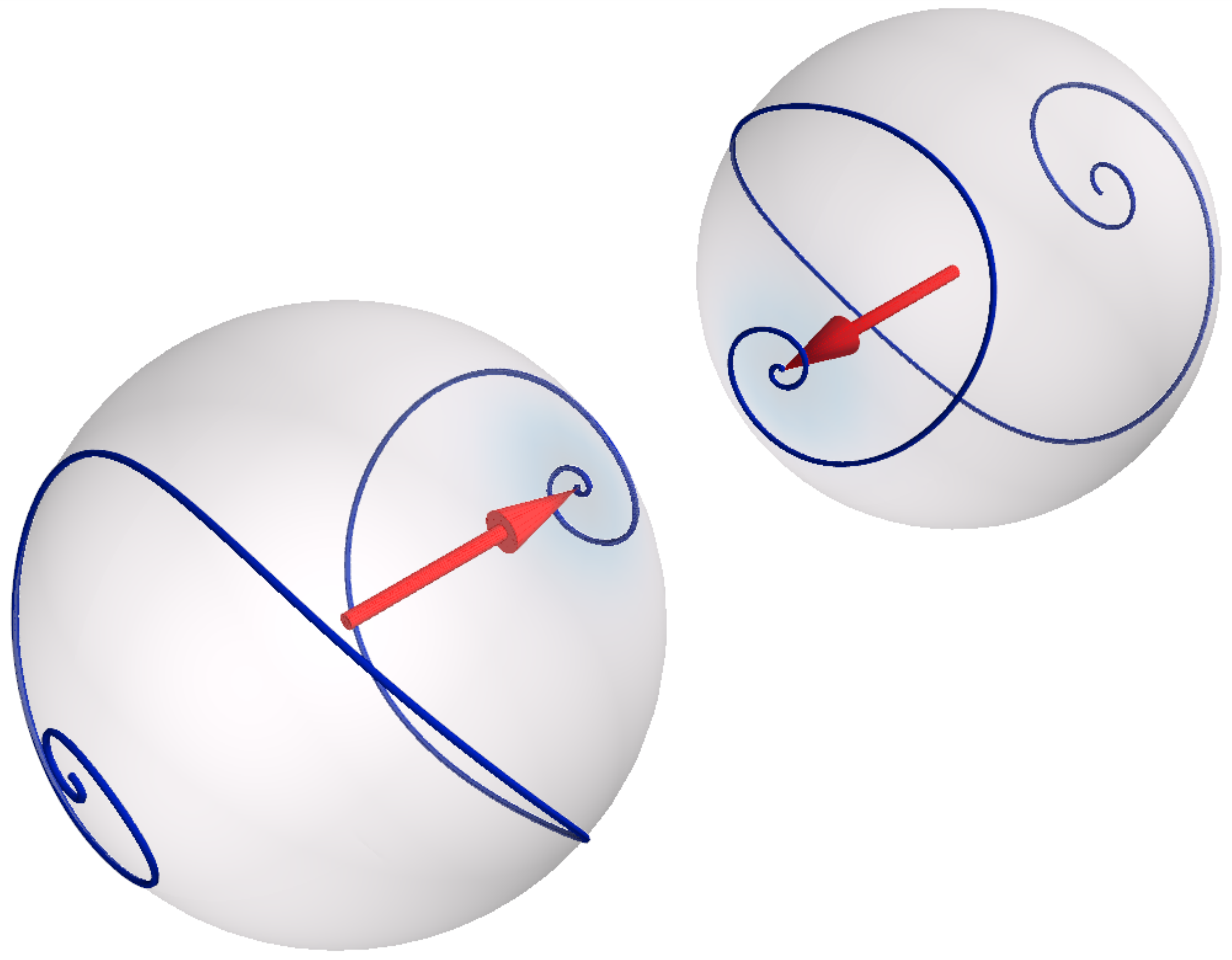}
        \caption{Classical motion of the initial state $\pmstate$ indicated
                by red arrows in the two Bloch spheres.
                A spiraling motion towards the anti-podal point
                leads to the spin-flipped state $\mpstate$.
                }
        \label{fig:classical_motion}
    \end{center}
\end{figure}
As a consequence, orbits in the neighborhood of the equilibrium are no
longer confined. In particular, an orbit reaching the equator of the Bloch
sphere spirals inwards to the anti-podal side, thus provides a classical
transfer, similar to the quantum transfer.
To obtain an analytical estimate of the classical transfer time
we model the quantum dynamics by classical orbits starting
a distance $\delta = \sqrt{\delta_1^2 + \delta_2^2}$ away
from an equilibrium point and determine the time $\tilde{t}_\text{cl}$
when they reach the equator at $x_1 = x_2 = 0$. By using the symmetry of the
transfer the total time is $2 \tilde{t}_\text{cl}$. On gets a logarithmic
dependence of the classical transfer time
\begin{equation}
 \tilde{t}_{\text{cl}}(\delta) =
    \frac{1}{c_1 c_2}
            \ln \left( \frac{\pi}{2\delta} \right), \\
\end{equation}
where
\begin{equation}
   c_1,c_2 = \sqrt{\frac{
    \sqrt{\coupling^2 + k_1 - k_2 - k_1 k_2 + 1}}{2}
    \mp \frac{(2 + k_1 - k_2)}{4}}.
\end{equation}
Figure~\ref{fig:fig2_arrival_time} demonstrates that this estimate gives
good agreement with the numerical results obtained from the classical ensemble
for various widths $\delta$ and different coupling $\coupling$.
\begin{figure}
    \begin{center}
        \includegraphics[width=8cm]{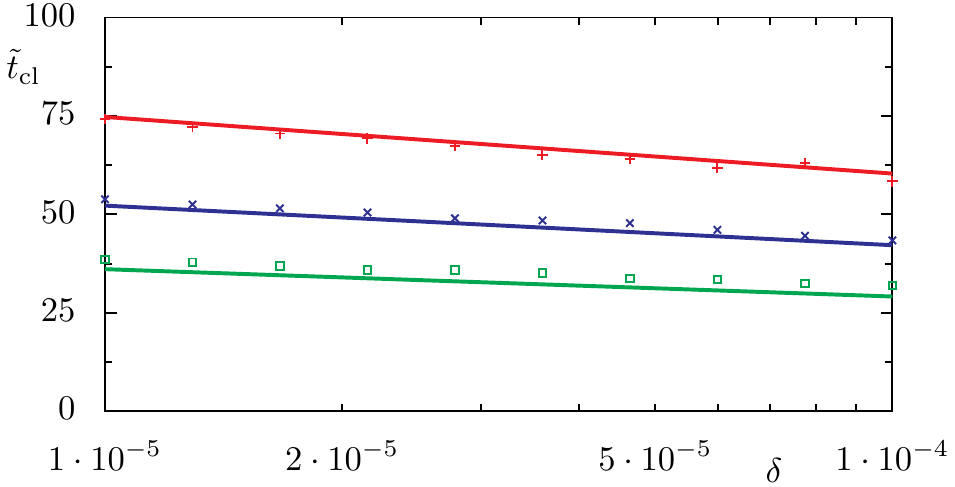}
        \caption{Numerically computed average time of an ensemble of orbits
which are a spherical distance $\delta$ on the Bloch sphere away from the
equilibrium point corresponding to $\pmstate$, to reach the equator of the
sphere. The solid lines show $\coupling = 1.05$, $1.1$, $1.2$
(red, blue, green, from top to bottom).}
        \label{fig:fig2_arrival_time}
    \end{center}
\end{figure}
For the final step towards a semiclassical description we consider the
quantum coherent states as a Gaussian distribution for each
subsystem, which is centered around the
equilibrium point with variance $\sigma^2 = \frac{2}{N}$. Hence we obtain after
averaging the initial conditions by such a radial normal
distribution
\begin{align}
 t_{\text{cl}} &= \frac{1}{\sigma^4}
        \int_0^\infty \ud r_1 \int_{0}^\infty \ud r_2 \, 2 r_1 r_2
                \tilde{t}_{\text{cl}}(x)
                \ue^{-(r_1^2 + r_2^2)/(2\sigma^2)}\\
    &= \frac{1}{c_1 c_2} \left(
                \ln(\frac{\pi^2}{8 \sigma^2}) + \gamma - 1
        \right)
\label{eq:trans_time_classic}
\end{align}

where $\gamma = 0.5772\ldots$ is the Euler-Mascheroni constant.

Using the classical transfer time \eqref{eq:trans_time_classic} allows for
establishing an approximate scaling relation of the quantum
transfer with respect to the system size $N = 2j + 1$ by using the re-scaled
time
\begin{equation} \label{eq:rescaled_time}
 \tau = \frac{t}{t_{\text{cl}}}.
\end{equation}
Thus for a fixed coupling $\coupling$ the transfer scales
similarly for different angular momenta $j$ in the re-scaled time, as
illustrated in Fig.~\ref{fig:fig_cu_transfer}, where also a parameter set is
depicted with different $k$ and $\epsilon$. This re-scaling works well only for
not to large differences of $\epsilon - k$.
Remarkably, the transfer scales logarithmic in $N$ and is thus in the same
order as the Ehrenfest time, which quantifies the time how long quantum
dynamics is expected to follow classical motion. Consequently the universal
transfer even works in the deep quantum regime of small sized spins,
$j \approx 5$.

Note that the change of the quantum transfer time is not abrupt at $\coupling =
\coupling_\crit$, but extends over the parameter interval $\epsilon_\crit \pm
1/N$, indicated by the vertical dashed red lines in
Fig.~\ref{fig:energy_level_plot}.

\begin{figure}
    \begin{center}
        \includegraphics[width=8cm]{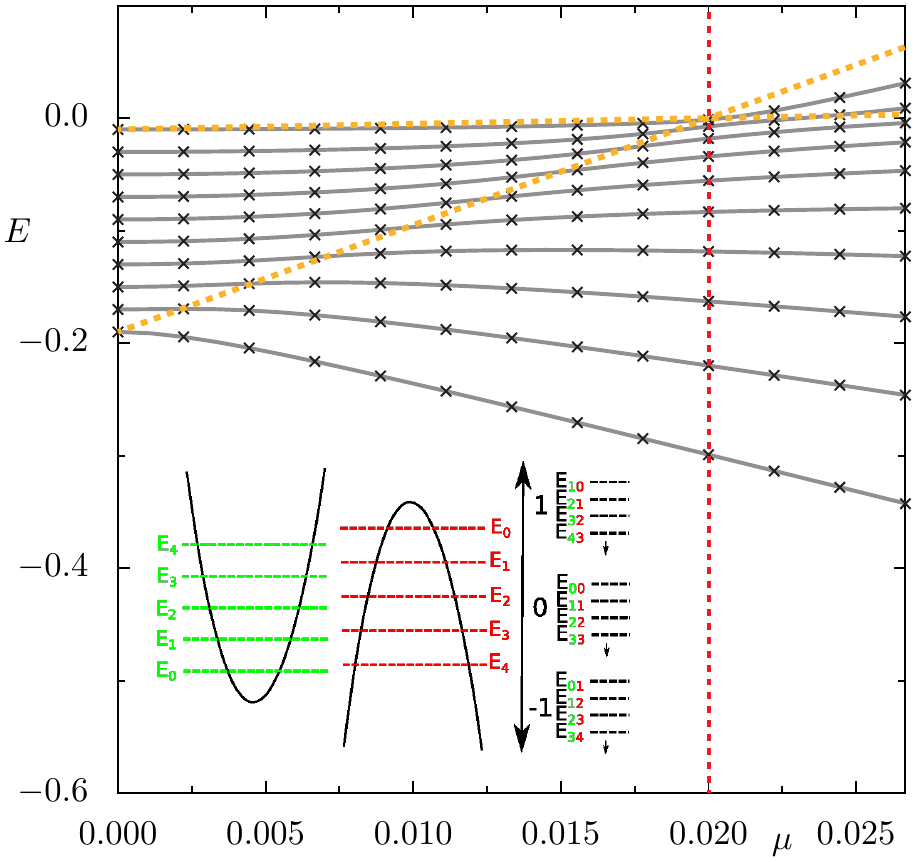}
        \caption{Energy levels of the quantized Hamiltonian
\eqref{eq:cherry_hamiltonian} for $M=10$ (crosses), compared to the
effective Hamiltonian \eqref{eq:effective_bundle_hamiltonian} (lines) for a
single cluster $\alpha = 0$ and $\Delta = 0.02$. The red dashed vertical line
indicates the EE-CU
transition, yellow dashed lines
the boundaries imposed by the Ger{\v s}gorin circle
theorem (only upper bounds). The inset illustrates how the
energy clusters arise for the uncoupled case.}
        \label{fig:cherry_hamiltonian_figures}
    \end{center}
\end{figure}

%%%%%%%%%%%%%%%%%%%%%%%%%%%%%%%%%%%%%%%%%%%%%%%%%%%%%%%%%%%%%%%%%%%%%%%%%%%%%
\emph{Model system with complex instability} --- The mechanism of the transfer
described for the system of two coupled angular momenta relies on the
possibility for complex instability but also on
the phase space organization. Thus in order to better understand the nature of
the EE to CU transition itself, we focus on the Cherry
Hamiltonian, which is locally similar to the coupled spin system. For this we
consider the semiclassical limit of the Hamiltonian
\eqref{eq:coupled_top_hamiltonian} by using canonical coordinates. By a Taylor
expansion of the Hamiltonian in one of
the equilibrium points, we obtain a system which locally reproduces the
elliptic-elliptic to complex unstable transition. Such a transition in its
simplest form occurs for the Cherry Hamiltonian which we use for further
analysis. The
classical Cherry Hamiltonian \cite{Che1928} describing two coupled harmonic
oscillators,
\begin{equation} \label{eq:cherry_hamiltonian}
 H_\text{cherry} = \frac{1}{2} \left( p_\suba^2 + q_\suba^2 \right)
        - \frac{w}{2} \left( p_\subb^2 + q_\subb^2 \right)
        + \cherrycoupling p_\suba p_\subb,
\end{equation}
where $w>0$.
The two coupled harmonic oscillators have opposite signs in their subsystems
energy. As a consequence the equilibrium at $(p_1, p_2, q_1, q_2) = (0, 0, 0,
0)$ exhibits a EE--CU transition for $\cherrycoupling_\crit = \frac{1 - w^2}{2
\sqrt{w}}$. The negative energy scale of one subsystem has important
consequences
on the energy level organization for which we consider the
case of small detuning of the oscillators, $\Delta := |w - 1| \ll 1$:
For an uncoupled system of finite subsystem size $M$ we find energy levels
the energy levels are $E_{n_\suba n_\subb} = n_\suba -  w n_\subb + 1/2 - w/2$
with $n_\ell = 0, \ldots, M-1$, see Fig.~\ref{fig:cherry_hamiltonian_figures}.
Thus the levels are given by clusters $\alpha = n_1 - n_2$ which are itself
similar to a single harmonic oscillator with frequency $\Delta$. Provided that
each cluster is well separated to the neighboring ones, only the level
interactions within a cluster are important. Hence we consider such a single
harmonic cluster described by an effective Hamiltonian
\begin{equation}
 \begin{aligned} \label{eq:effective_bundle_hamiltonian}
 H_\text{cluster} &= -\Delta \sum_{m = 1}^{M} m \dyad{m}\\
        &+ \sum_{m = 2}^{M} \frac{\cherrycoupling m}{2}
            \left( \dyad{m}{m-1} + \dyad{m-1}{m} \right).
 \end{aligned}
\end{equation}
This reproduces the essential behavior of the Cherry Hamiltonian at the EE--CU
transition, see Fig.~\ref{fig:cherry_hamiltonian_figures}. In the basis $\{
\ket{m} \}$ the
Hamiltonian is represented by a tri-diagonal matrix.
The Ger{\v s}gorin circle theorem
\cite{Ger1931}, simplified to this special case,
Eq.~\eqref{eq:effective_bundle_hamiltonian}, states that all
eigenvalues have to lie within the union of intervals with lengths given by the
sum of the off-diagonal matrix-rows, centered around the value of the diagonal
entry. This provides boundaries to the range of the eigenvalues.
We observe that the upper constraint from the lowest and the highest energy
cross at $\mu = \Delta$, see the yellow lines in
Fig.~\ref{fig:cherry_hamiltonian_figures}, which is close to the transition
point of the classical system, leading to a larger range of the eigenvalues in
the CU regime. The crossing in which the highest and lowest eigenvalue are
connected, already indicates a transition
and thus establishes some kind of long-ranged correlation.
This becomes more explicit by considering the ``ground
state'' of the system, which is the eigenstate with largest eigenvalue.
We find that this state is exponentially localized in the EE regime and
becomes delocalized at the transition point (not shown).
In the EE regime the state can be described as an exponentially localized vector
$\ket{\Psi} = \sum c_j \ket{\Psi_j}$ with $c_j \sim
\ue^{-j/l}$ and localization length $l$. Using this ansatz in
Eq.~\eqref{eq:effective_bundle_hamiltonian} leads to
a self-consistent expression for $l$, which diverges at the critical point
\begin{equation} \label{eq:localization_length}
 l = \frac{1}{\ln \left( \frac{\Delta}{\cherrycoupling} +
\sqrt{\left(\frac{\Delta}{\cherrycoupling}\right)^2 - 1} \right)}.
\end{equation}
This diverging localization length results in a delocalized eigenfunction.
This delocalization in the simplified system can be seen as the corresponding
effect for the spin system \eqref{eq:coupled_top_hamiltonian}, where the
initial state becomes distributed over multiple eigenstates.
The classical interpretation of this effect
can be understood by the spiraling motion of orbits,
which are no longer confined and explore a large region of phase space.

%%%%%%%%%%%%%%%%%%%%%%%%%%%%%%%%%%%%%%%%%%%%%%%%%%%%%%%%%%%%%%%%%%%%%%%%%%%%%
\emph{Summary and outlook} \label{seq:summary_outlook} --- For two
opposing equilibrium points of the proposed bipartite system
\eqref{eq:coupled_top_hamiltonian}
changing the coupling induces a transition from elliptic-elliptic stability
to complex unstable dynamics in the semiclassical limit.
The counter-rotating, bipartite nature of the initial state configuration and
the geometry of phase space allows for a fast and energy-efficient transfer
from the state $\pmstate$ into the bit-flipped state $\mpstate$ compared to 
dynamical tunneling.
These results can also be extended to the case of a time-periodic kicked
system, namely the coupled kicked tops for which we obtain qualitatively
similar results.
Of future interest is the situation of parameters which are directly on
the transition boundary, where a power-law dependence of the transition time
is expected.
We propose an experimental investigation of the system, either as a kicked or
as an autonomous realization.
Particularly interesting would be to extend the system to multiple bits,
which would allow for investigating the emergence of complex instability
in many-body systems,
such as in Bose-Hubbard systems which have a semiclassical limit.

%%%%%%%%%%%%%%%%%%%%%%%%%%%%%%%%%%%%%%%%%%%%%%%%%%%%%%%%%%%%%%%%%%%%%%%%%%%%%
\acknowledgments

We thank Tabea Herrmann, Roland Ketzmerick, and Jan Robert Schmidt
for useful discussions.
Funded
by the Deutsche Forschungsgemeinschaft (DFG, German
Research Foundation) – 290128388; 497038782

\bibliographystyle{cpg_unsrt_title_for_phys_rev.bst}

\end{document}